\documentstyle[11pt]{article}
\input epsf

\begin{document}

\baselineskip 24pt

\title{\Large\bf
 Phenomenology of the Top Mass in Realistic Extended Technicolor Models}

\author{Thomas Appelquist\thanks{Electronic address:
 twa@genesis3.physics.yale.edu},
 Nick Evans\thanks{Electronic address:
 nick@zen.physics.yale.edu}\ \ and
 Stephen B. Selipsky\thanks{Electronic address:
 stephen@genesis1.physics.yale.edu}\\ \\ {\it
 Department of Physics, Yale University, New Haven, Connecticut 06520-8120}}

\date{January 21, 1996}

\maketitle

\begin{picture}(0,0)(0,0)
\put(365,245){hep-ph/9601305}
\put(365,230){YCTP-P1-96}
\end{picture}
\vspace{-24pt}

\begin{abstract}
   Extended technicolor (ETC) theories typically require ETC gauge bosons
lighter than of order 1 TeV, to perturbatively generate the $t$ quark mass.
We point out that explicit models of $t-b$ mass splitting also typically
contain additional TeV scale ETC gauge bosons transforming in the
{\it adjoint} of technicolor, leading to large weak-isospin-breaking
effects observable in the $\rho$ parameter.
Viable ETC models may thus require a lowest ETC scale of order 10 TeV,
with relatively strong and finely tuned couplings to generate $m_t$.
Such models do not generate observable corrections to the $Zb{\bar b}$
vertex.
\end{abstract}

   Technicolor models break electroweak symmetry with fermion condensates,
generated by a strongly interacting gauge theory patterned after QCD
\cite{TC}.  The masses and mixing angles of quarks and leptons arise from
an extended technicolor (ETC) sector \cite{ETC} that communicates the
spontaneous breaking of electroweak symmetry to the quark-lepton sector
through broken gauge interactions.  To accommodate a top mass
$m_t \simeq 175$ GeV with perturbative ETC interactions requires a
corresponding ETC scale less than or on the order of 1 TeV.

   Such a low scale for new physics raises the possibility that ETC dynamics
may visibly affect low energy precision data; in particular there has
recently been considerable interest \cite{zbbtheory,diagETCrho} in relating
ETC dynamics to shifts observed \cite{lepwg} in the $Zb{\bar b}$ width.
In addition, the weak interaction $\rho$ parameter can receive important
corrections from ETC interactions \cite{tomdrho} which exhibit enough
weak-isospin-breaking at the lowest ETC scale to explain the mass
splitting between the top quark $t$ and the bottom quark $b$.
Such a large splitting requires realistic ETC models to be chiral
\cite{ETC, chiralETC}, treating the $t_R$ and $b_R$ differently
yet allowing the $t_L$ and $b_L$ to transform together.

   We shall consider chiral models with separate ETC groups acting on the
$t_R$ and $b_R$, allowing separate ETC couplings and breaking scales to be
associated with each.  In order to most naturally and simply accommodate
the three generation structure of the standard model, we assume that the
ETC gauge group commutes with $SU(2)_L$ and that the fermions transform
in the fundamental representation of ETC.  Alternative implementations of
chirality might put the $t_R$ and $b_R$ into different representations
of a single ETC group \cite{ETC,repsterning}, and even bring $SU(2)_L$
elements into the ETC group (``non-commuting'' ETC models \cite{noncommute}).
However, such models require a constrained choice of representations
to avoid ending up with technifermions in different representations of
technicolor, and hence forcing the technicolor interactions to directly
violate weak-isospin symmetry.  In the absence of complete models of this
sort, we shall concentrate on the multi-group case.

   The massive ETC gauge bosons in such models have couplings which
violate weak-isospin symmetry and hence contribute to
 $\Delta\rho \equiv \rho - \rho^{\rm (SM)}$,
or $\alpha T$ in the notation of Ref.~\cite{pesktak}.
The contributions from ETC bosons associated with the broken
{\it diagonal} generators of the ETC gauge groups have previously been
calculated \cite{diagETCrho} to be near experimental limits.
In this paper we point out that these models also contain massive ETC
bosons in the {\it adjoint} representation of technicolor.  Their exchange
gives rise to $\Delta\rho$ contributions exceeding the experimental limits
by at least an order of magnitude, if the corresponding $M_{ETC}$ scales
are of order 1 TeV (small enough to generate $m_t$ perturbatively).
Alternatively, if ETC bosons are an order of magnitude heavier,
to adequately suppress the adjoint contribution to $\Delta\rho$, then
generating $m_t$ requires strong ETC interactions \cite{strongETC}.
Furthermore, such heavy ETC bosons do not observably correct the
$Zb{\bar b}$ vertex.

   At the lowest ETC scale, the ETC gauge groups break to a single
asymptotically free technicolor group, with $N_D$ electroweak doublets
of technifermions transforming in some representation of technicolor.
At the scale $\Lambda_{TC}$, the technicolor group confines technifermions
and breaks electroweak symmetry.
The electroweak scale $v = 250$ GeV can be related directly to $\Pi_{AA}$,
the coefficient of $g_{\mu\nu}$ in the axial-axial current correlator
that generates $m_W$ and $m_Z$.
Summing over the contributions to $\Pi_{AA}$ from single loops of each
technifermion flavor $n$,
\begin{equation} \label{vsumpiaa}
 v^2 \ \simeq\ {1\over 4} \sum_{n}^{}{\Pi_{AA}^{(n)}(p^2 = 0)} \ .
\end{equation}
We can crudely estimate the correlators, to zeroth order in ETC interactions,
by assuming that a Pagels--Stokar formula \cite{PagelsStokar} with dynamical
masses $\Sigma(k)$ adequately approximates the technifermion dynamics.
Taking $\Sigma(0) \simeq \Lambda_{TC}$ gives
\begin{equation}
 v^2 \ \simeq\ {D(R) N_D \over 2} \int{dk^2 \over 2\pi^2}
 {k^2 \, \Sigma^2(k) \over (k^2 + \Sigma^2(k))^2} \ \simeq\
 D(R) N_D {\Lambda_{TC}^2 \over 4 \pi^2} \ ,
\end{equation}
where $D(R)$ is the dimensionality of the technifermion representation.
In our subsequent discussion we shall assume for simplicity that
technifermions occupy the fundamental representation of technicolor,
for which $D(R) = N_{TC}$.  If the technifermions are in larger
representations, then the group theory factors are larger
and the phenomenological difficulties even worse.

   The $t$ and $b$ masses must be generated by ETC interactions connecting
$t$ and $b$ quarks to their respective technifermion partners.
In accordance with our earlier discussion, we assume that above the lowest
ETC scale there are at least two groups, one acting on $t_R$ and one on $b_R$,
whose product contains technicolor.
As specific examples of multi-group ETC models, let us examine the ETC
breaking patterns responsible for the $t$ and $b$ masses, in a one doublet
technicolor model and in a one family technicolor model.
We concentrate on the symmetry breaking patterns rather than the breaking
mechanisms, simply noting that among other possibilities an underlying
QCD-like model \cite{chiralmoose} can trigger the necessary breaking.
Whatever the breaking mechanism and technifermion content, the models
should respect the stringent constraints of precision electroweak data,
including the $S$ parameter measurements \cite{STfit} which favor models
with fewer technifermions.

   In a one doublet technicolor model, the ETC fermion multiplets above the
lowest ETC scale are ${\cal U}_R$, ${\cal D}_R$ and $({\cal U},{\cal D})_L$,
which contain the technicolor multiplets $(U,D)$ and the QCD triplets
of $t$ and $b$ quarks \cite{PatiSalam}.
A particularly simple example of a chiral gauge structure puts both
$({\cal U},{\cal D})_L$ and ${\cal U}_R$ into the same (fundamental)
representation of a single ETC subgroup $SU(N+3)_L$,
while ${\cal D}_R$ transforms under a separate group $SU(N+3)_{{\cal D}_R}$.
At the ETC-breaking scale, the group structure then breaks in the pattern
\begin{equation} \label{onedoublet}
 SU(N+3)_L \times SU(N+3)_{{\cal D}_R} \rightarrow
 SU(N)_{TC} \times SU(3)_{QCD} \ .
\end{equation}
We may wish to distinguish between possibly separate breaking scales:
$F_L$, of order 1 TeV, for $SU(N+3)_L \rightarrow SU(N)_L \times SU(3)_L$;
$F_{{\cal D}_R}$, of order 1 TeV or larger, for $SU(N+3)_{{\cal D}_R}$
similarly;
and $F_{\rm mix}$, less than or of order 1 TeV, for the ``vector subgroup''
mixing.

   In a one family technicolor model, the ETC fermion multiplets must
contain at least the technicolor multiplets $(U,D)$ and $(N,E)$ and
the full third generation of standard fermions.
Above the lowest ETC scale, these ETC multiplets include
${\cal U}_R$, ${\cal D}_R$, ${\cal N}_R$, ${\cal E}_R$,
$({\cal U, D})_L$ and $({\cal N, E})_L$.
An especially simple gauge structure, as above, is
$SU(N+1)_L \times SU(N+1)_{{\cal D}_R}$,
where ${\cal D}_R$ transforms under $SU(N+1)_{{\cal D}_R}$
and all the other multiplets under $SU(N+1)_L$.  The breaking pattern,
\begin{equation} \label{onefamily}
 SU(N+1)_L \times SU(N+1)_{{\cal D}_R} \rightarrow SU(N)_{TC} \ ,
\end{equation}
again takes effect at scale(s) $F_L, F_{{\cal D}_R}$ and $F_{\rm mix}$.

   Independently of these particular models, in general each ETC group
contains bosons transforming in the adjoint, the fundamental, and the
singlet representations of technicolor.  Since only one adjoint remains
massless below the ETC scales to form the technicolor gauge bosons,
chiral ETC models generate many massive ETC bosons.
The group structure could be more elaborate than in the simple examples above:
each right-handed ETC multiplet might transform under a separate ETC group,
and a sufficiently grandiose model could also distinguish different left-handed
multiplets.  Enlarging the number of distinct ETC groups simply creates more
sets of massive gauge bosons, exacerbating the phenomenological difficulties
described below.

\begin{figure}[hbt]
\hfil{
 \epsfysize = 1.5in
 \epsfbox{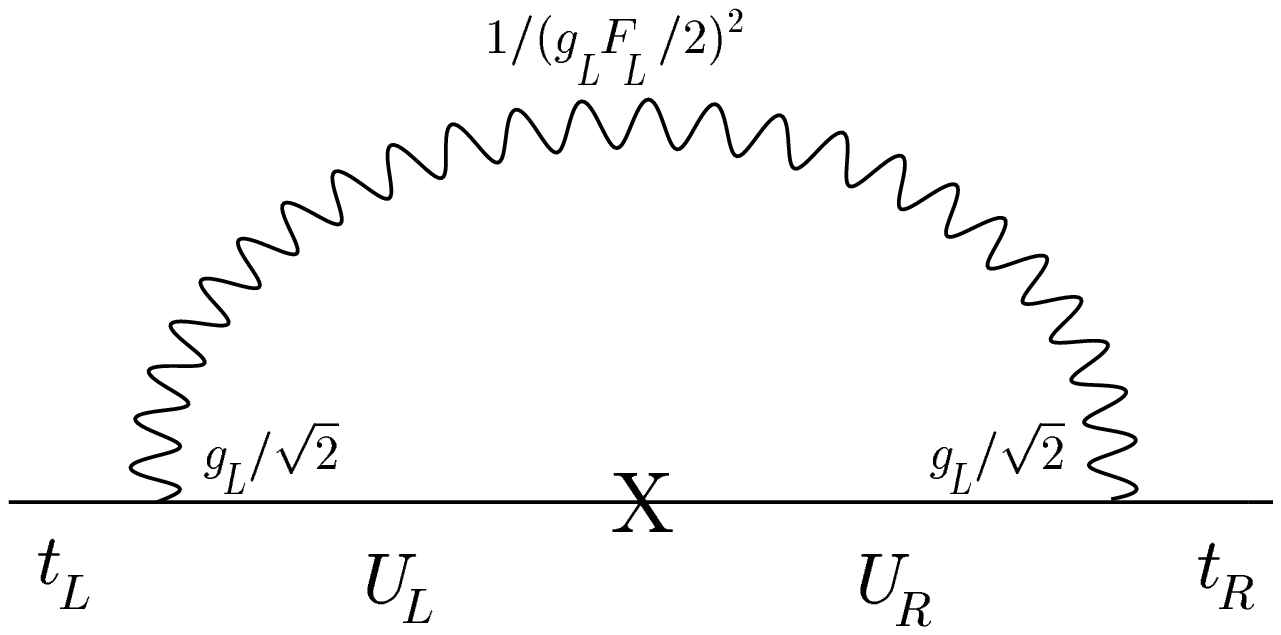}
 \hfil
 \epsfysize = 1.75in
 \epsfbox{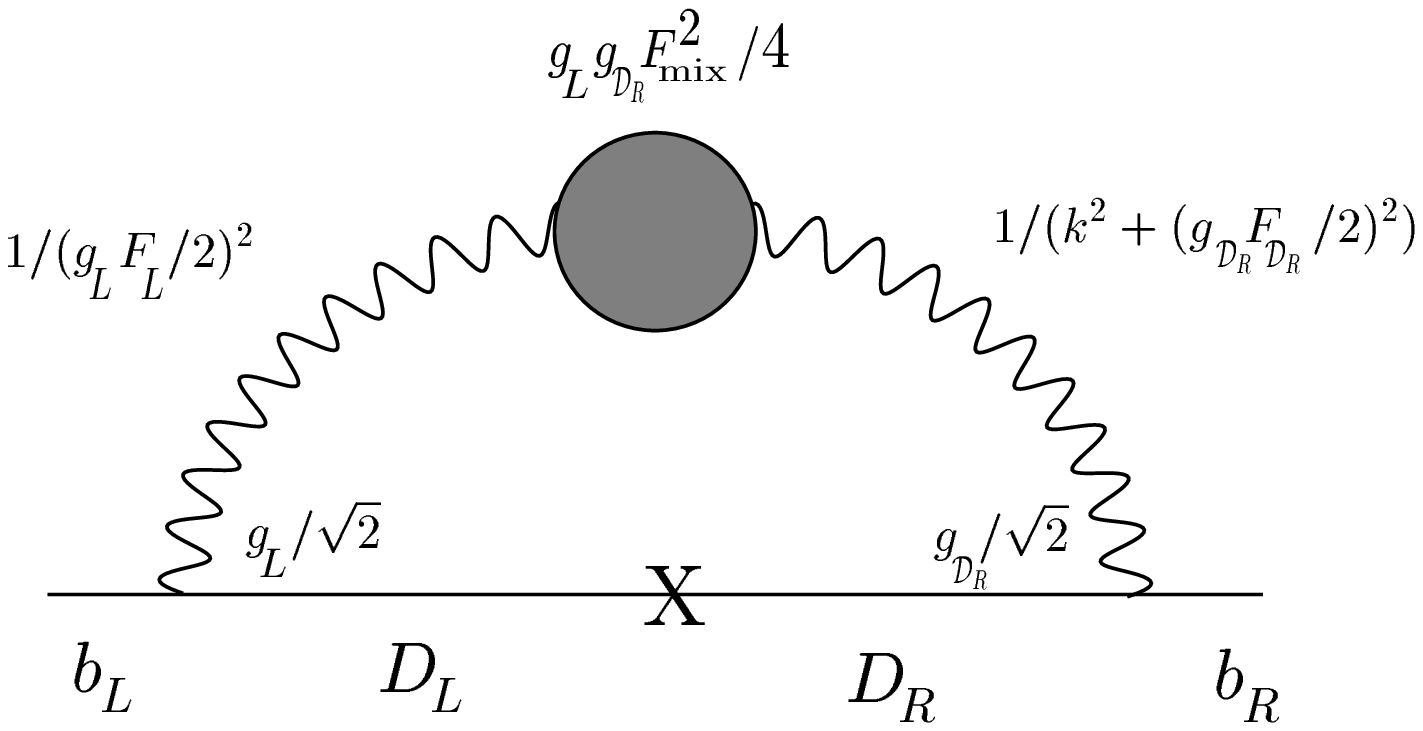}
 }\hfil \vspace{-0.45in}
\caption{
 Perturbative generation of top and bottom masses, schematically indicating
 some coupling constant factors and heavy boson propagators.
 The heavy X represents a technifermion condensate, the blob represents
 ETC boson mixing induced by breaking to the ``vector subgroup''.
 The ${\cal D}_R$ boson requires a full propagator if it is light.
} \label{fig1}
\end{figure}

   The $t$ and $b$ masses are generated by the ETC bosons transforming in
the fundamental representation of technicolor (``sideways'' ETC bosons).
There is one set of such bosons associated with each ETC gauge group,
and they generate $m_t$ and $m_b$ as shown in Fig.~\ref{fig1}.
The ETC couplings cancel in the four-fermion approximation, which is
applicable if the ETC boson masses ($\simeq gF/2$, with $g$ the appropriate
ETC coupling) are at least of order 1 TeV, larger than the dominant
internal momentum (the technicolor scale) in the diagram.
Then from the first diagram, $m_t \simeq \langle {\bar U}U \rangle / F_L^2$,
where $F_L$ is the decay constant of the Goldstone bosons formed and
eaten at the breaking scale, and where the techni-up condensate
$\langle {\bar U}U \rangle$ is roughly of order $4\pi v^3$.
Generating $m_t \simeq 175$ GeV requires $F_L
 {\ \lower-1.2pt\vbox{\hbox{\rlap{$<$}\lower5pt\vbox{\hbox{$\sim$}}}}\ }
1$ TeV.

   The second diagram similarly generates $m_b$, with additional factors
from the ETC boson mixing and extra ETC (${{\cal D}_R}$) boson propagator.
If the latter boson's mass, approximately $g_{{\cal D}_R} F_{{\cal D}_R}/2$,
is large enough to allow the four-fermion approximation,
then $F_{{\cal D}_R}$ must exceed the other breaking scales so that
$F_{\rm mix}^2 / F_{{\cal D}_R}^2$ suppresses $m_b$ relative to $m_t$.
Alternatively, if $F_L \simeq F_{\rm mix} \simeq F_{{\cal D}_R}$,
then obtaining the necessary $t-b$ mass splitting requires
$g_{{\cal D}_R}$ to be much less than unity.
The associated ETC boson is in that case lighter than the momentum scale in
the diagram, invalidating the four-fermion approximation; $g_{{\cal D}_R}^2$
then remains uncancelled in the numerator, suppressing $m_b$.

   The small coupling alternative faces a potential problem.
The technicolor coupling must run from its value at the lowest ETC
breaking scale, where it is a function of the ETC couplings and no larger
than the smallest of them, to its confining value at the technicolor scale.
If $m_b$ is the result of a large value for $F_{{\cal D}_R}$,
this should not cause any difficulty since all the ETC couplings can be
at least of order unity at the lowest ETC scale (although below $2\pi$
if the ETC interactions are perturbative).
However, if all the ETC scales are below of order 1 TeV, then as noted
above the smallness of $m_b$ is due to the smallness of $g_{{\cal D}_R}$.
The technicolor coupling at this scale must then also be small and might
have insufficient range to run to its confining strength at the weak scale.
Despite this potential problem, we will continue to consider the small
$g_{{\cal D}_R}$ alternative as we explore the phenomenology of this
class of models.

   To make more precise estimates of $m_t$ and $m_b$, we work with
the gauge boson mass eigenstates of each technicolor representation.
Mixing angles generically of order unity, generated by mass mixing at
the ETC breaking scales, relate these mass eigenstates to the ETC gauge
eigenstates associated with ${{\cal U}_R}$, ${{\cal D}_R}$,
and possible additional bosons in more elaborate models.
When $m_b$ is the result of a small coupling $g_{{\cal D}_R}$, a light
mass eigenstate occurs in the technicolor fundamental representation,
coupled mainly to the $b$.  Similarly one other eigenstate, with mass
at least of order 1 TeV, couples mainly to the $t$.
When $m_b$ is instead the result of a large ETC scale $F_{{\cal D}_R}$,
the mainly $b$-coupled ETC mass eigenstate is heavier than both the
technicolor scale and the $t$-coupled state.
The mixing angles are in this case functions of only the breaking scales,
since in the four-fermion approximation the couplings cancel as discussed
above.  In either case, we sum the contributions to $m_t$ and $m_b$
generated by each mass eigenstate.

   Such contributions from ETC bosons heavier than the technicolor scale
take the form
\begin{eqnarray} \label{topmassETC}
 m_f^{\rm (heavy\ ETC)} & \simeq & \int{dk^2 \over 4\pi^2}
 N_{TC} \left( (V_{fF}^{\rm fund})^2 - (A_{fF}^{\rm fund})^2 \right)
 {k^2\over M^2_{\rm fund}} {\Sigma(k)\over k^2 + \Sigma^2(k)} \nonumber\\
 & \simeq & N_{TC} \left( (V_{fF}^{\rm fund})^2 - (A_{fF}^{\rm fund})^2 \right)
  {\Lambda_{TC}^3\over 4\pi^2 M^2_{\rm fund}}\ \\
 & \simeq & \left((V_{fF}^{\rm fund})^2 - (A_{fF}^{\rm fund})^2\right)
 {2 \pi v^3\over \sqrt{N_{TC}N_D^{3}}\, M^2_{\rm fund}}\ .\nonumber
\end{eqnarray}
Here $V_{fF}^{\rm fund}$ and $A_{fF}^{\rm fund}$ are the vector and axial
vector couplings, to the fermion-technifermion pair $(f,F)$, of the fundamental
representation ETC boson mass eigenstates with mass $M_{\rm fund}$.
Up to group theory factors and mixing angles, these couplings are
combinations of $g_L/\sqrt{2}$ and $g_{{\cal D}_R}/\sqrt{2}$.
In fact, with perturbative ETC interactions the couplings cancel between
$V_{fF}$ or $A_{fF}$ and $M_{\rm fund}$, giving from this expression a
rough estimate $m_t \simeq 4\pi v^3 /(F_L^2 \sqrt{N_{TC} N_D^3})$.
To make this contribution 175 GeV requires $F_L$ just below 1 TeV,
even in the minimal model with $N_{TC} = 2$, $N_D = 1$.
If $m_b$ is small as the result of a large ETC scale instead of a small
coupling, we similarly obtain
 $m_b \simeq \left(F_{\rm mix} / F_{{\cal D}_R}\right)^2 m_t$.
This is sufficiently suppressed if $F_{{\cal D}_R} \simeq 5$ TeV.

   In the small coupling case, however, ETC bosons can exist below the
technicolor scale.  Their contributions to $m_t$ and $m_b$ take the form
\begin{eqnarray} \label{topmasslightETC}
 m_f^{\rm (light\ ETC)} &\ \simeq\ & \int {dk^2 \over 4\pi^2}
  N_{TC} \left( (V_{fF}^{\rm fund})^2 - (A_{fF}^{\rm fund})^2 \right)
  {\Sigma(0)\over k^2 + \Sigma^2(k)} \nonumber\\
 & \simeq & N_{TC} \left( (V_{fF}^{\rm fund})^2 -
  (A_{fF}^{\rm fund})^2 \right) {\Lambda_{TC}\over 4\pi^2} \\
 & \simeq & {\left( (V_{fF}^{\rm fund})^2 - (A_{fF}^{\rm fund})^2 \right)
  \over 2 \pi} \sqrt{ N_{TC} \over N_D } v \ .\nonumber
\end{eqnarray}
To obtain $m_b \simeq 5$ GeV in this small coupling regime, we estimate
$\left( (V_{bD}^{\rm fund})^2 - (A_{bD}^{\rm fund})^2 \right) \sim
 g_{{\cal D}_R}^2 \simeq (0.2)^2$, which can dominate the contributions
to $m_b$.  The weak coupling responsible for the light gauge boson mass
leaves the corresponding contribution to $m_t$ unimportant relative to
the heavy-boson contributions.

   We turn next to implications of chiral ETC for low energy precision
measurements, estimating first the contributions to $\Delta\rho$.
The dominant contributions are from loops of technifermions corrected
by exchange of ETC bosons transforming in the adjoint representation
of technicolor.
Recall that with at least two ETC groups at the lowest ETC scale,
there are at least two sets of gauge bosons transforming in the adjoint
representation of technicolor.
One set, the technigluons of unbroken technicolor, remains massless,
with gauge coupling smaller than the smallest ETC coupling.
The other set (with coupling of order the largest ETC coupling)
acquires a mass of order at least 1 TeV set by the scale $F_{\rm mix}$,
independently of whether $g_{{\cal D}_R}$ is small or $F_{{\cal D}_R}$
large.
Within each ETC group, the gauge bosons in different representations of
technicolor share the same ETC coupling and breaking scale,
which we have related to the $t-b$ mass splitting.
The couplings of the massive adjoints, therefore, explicitly violate
weak-isospin symmetry.  Technicolor-singlet ``diagonal'' bosons generally
also exist, with masses equal to the fundamentals discussed previously.

\begin{figure}[hbt]
\hfil{
 \epsfysize = 1in
 \epsfbox{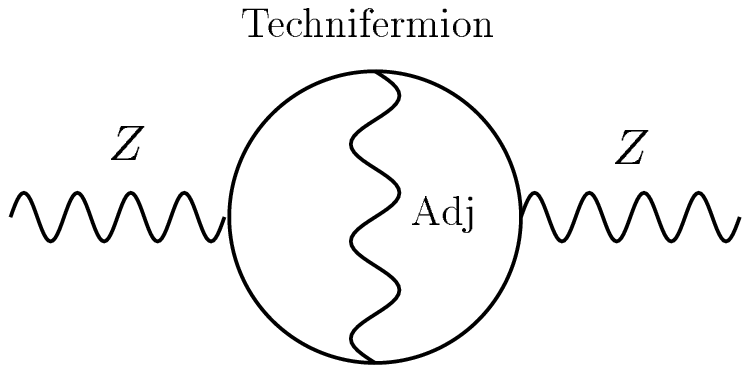}
 \hfil
 \epsfysize = 1.2in
 \epsfbox{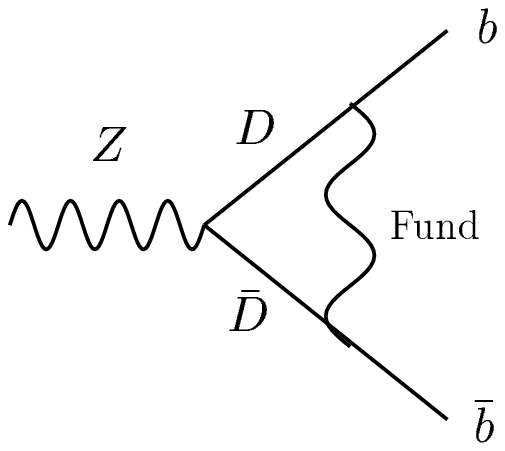}
 \hfil
 \epsfysize = 1in
 \epsfbox{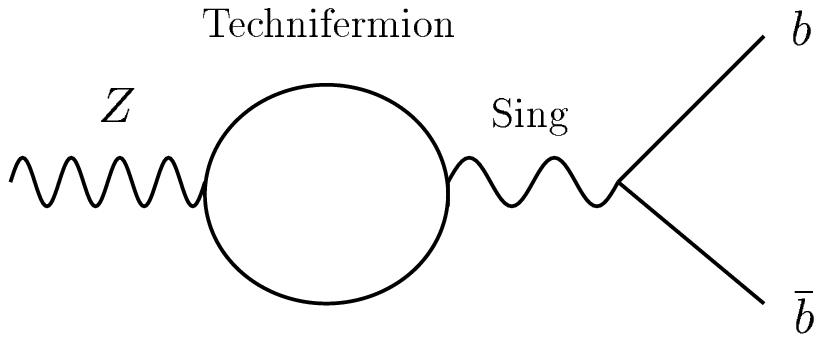}
 }\hfil
\caption{
 (a): Contribution to $m_Z$ from an ETC boson in the technicolor adjoint
      representation, exchanged across a technifermion loop.
 (b): $Zb{\bar b}$ vertex correction from a ``sideways'' ETC boson in the
      technicolor fundamental representation.
 (c): $Zb{\bar b}$ vertex correction from a ``diagonal'' ETC boson in the
      technicolor singlet representation, with which the $Z$ has mixed.
} \label{fig2}
\end{figure}

   Let us first consider the lowest order contribution shown in
Fig.~\ref{fig2}a.  Since the adjoint representation ETC gauge boson mass,
$M_{\rm adj}$, exceeds the technicolor scale we simply generate
an effective four-fermion interaction, and after a Fierz transformation
the diagram becomes the product of two axial-axial current correlators.
Neglecting strong technicolor interactions between the two technifermion
loops, the contribution to the $Z$ mass becomes
\begin{equation}
 \Delta M_Z^2 \ \simeq\ -{e^2 \over 4 s_{\theta}^2 c_{\theta}^2}
 {v^4 \over N_D^2 M_{\rm adj}^2}
 \left[ (A_U^{\rm adj})^2 + (A_D^{\rm adj})^2 \right]
\end{equation}
where the $A_n^{\rm adj}$ are the axial couplings of the ETC bosons
to technifermion flavor $n$.  The $W$ mass is similarly corrected by
\begin{equation}
 \Delta M_W^2 \ \simeq\ -{e^2 \over 2 s_\theta^2}
 {v^4 \over N_D^2 M^2_{\rm adj}} A_U^{\rm adj} A_D^{\rm adj} \ .
\end{equation}
Although for convenience we have expressed the results in terms of the
ETC couplings and $M_{\rm adj}$, as discussed previously the ETC couplings
approximately cancel at lowest order.

   The lowest order contribution to
$\Delta\rho \equiv M_W^2/c^2_\theta M_Z^2 - 1$ is thus
\begin{equation} \label{rho1adj}
 \Delta\rho_{(1)} \ \simeq\ {v^2 \over N_D^2 M_{\rm adj}^2}
 \left( A_U^{\rm adj} - A_D^{\rm adj} \right)^2 \ .
\end{equation}
In the models of both Eqs.~(\ref{onedoublet}) and (\ref{onefamily}),
the ETC axial couplings to the $U$ vanish, but $A_D^{\rm adj}$
is of order the largest of $g_L$ and $g_{{\cal D}_R}$,
independent of small $g_{{\cal D}_R}$ or large $F_{{\cal D}_R}$.
(In more complicated models, $(A_U^{\rm adj} - A_D^{\rm adj})$
remains non-zero, reflecting the isospin splitting that must be present
to generate $t-b$ mass splitting).
Taking $(A_D^{\rm adj})^2/M_{\rm adj}^2 \simeq (1\ {\rm TeV})^{-2}$
we find $\Delta\rho_{(1)}
{\ \lower-1.2pt\vbox{\hbox{\rlap{$>$}\lower5pt\vbox{\hbox{$\sim$}}}}\ }
 6\%$.  Even if trusted only in order of magnitude given the approximate
treatment of strong technigluon interactions in Fig.~\ref{fig2}a,
this is completely at odds with experiment.

   Additional contributions to $\Delta\rho$ arise at first order.
Each ETC group also contains a massive ETC boson in the technicolor
singlet representation, whose contributions to $\Delta\rho$ have been
calculated \cite{diagETCrho} to be of order $1\%$ or less.
There may be other contributions, from ETC bosons that do not couple
to the technifermions (for example, bosons in the adjoint of QCD,
in a one doublet model).
Since these contributions arise in loops of $t$ and $b$ quarks,
they are suppressed by at least $(m_t/\Lambda_{TC})^2$.
Clearly, for $M_{\rm adj}$ of order 1 TeV, the adjoint ETC bosons
dominate the contributions to $\Delta\rho$.

   Raising the lowest ETC scale above about 10 TeV quadratically suppresses
the first order contribution $\Delta\rho_{(1)}$ to below current experimental
bounds.  Simultaneously maintaining $m_t \simeq 175$ GeV despite this
decoupling, however, then requires tuning the ETC coupling $g_L$ close to
its critical value.  The degree of tuning is at least of order $10\%$ for
a 10 TeV ETC scale \cite{strongETC}.

   Maintaining $m_t = 175$ GeV, though, prevents decoupling of some
higher order contributions to $\Delta\rho$, due to technifermion $U-D$
mass splitting arising at second order in ETC boson exchange.
When the splitting is small relative to the technifermion masses
this ``indirect'' contribution \cite{pesktak} yields approximately
\begin{equation} \label{rho2}
 \Delta\rho_{(2)}\ \simeq\
 {N_{TC} \over 12\pi^2 v^2} \left[ \Delta\Sigma(0) \right]^2 \ \simeq\
 0.4\% \times N_{TC} \left( {\Delta\Sigma \over m_t} \right)^2 \ .
\end{equation}
We expect $\Delta\Sigma(M_{ETC})$ to be of order $m_t$, since the $t$ unifies
with the $U$ (and similarly the $b$ with the $D$) at the lowest ETC scale,
where their masses must be equal.
That equality is preserved as the ETC scale increases.
The massive adjoint gauge bosons generate the dominant corrections to the
mass of the $i^{\rm th}$ technifermion flavor, calculated similarly to the
quark masses:
\begin{eqnarray} \label{UDsplitting}
 \Delta\Sigma_i & \ \simeq\ & \int d^2k {N_{TC}\over 4\pi^2}
 \left( (V_{F_i}^{\rm adj})^2 - (A_{F_i}^{\rm adj})^2 \right)
 {k^2\over M^2_{\rm adj}} {\Sigma_i(k)\over k^2 + \Sigma_i^2(k)}
\end{eqnarray}
in the four-fermion approximation, leaving $\Delta\Sigma_i$ almost
scale-independent.
This is identical to Eq.~(\ref{topmassETC}), the main contribution to $m_t$,
except for different $V$ and $A$ couplings that reflect the different
diagonalization of the adjoint instead of fundamental mass matrices.
Thus if critical behavior enhances $m_t \simeq 175$ GeV
then it should similarly enhance $\Delta\Sigma_U$.

   Eqs.~(\ref{rho2}) and (\ref{UDsplitting}) show that when the ETC couplings
are perturbative, $\Delta\rho_{(2)}$ is an order of magnitude smaller than
$\Delta\rho_{(1)}$; but that when the ETC breaking scale is raised with the
ETC coupling tuned to maintain $m_t$,
$\Delta\rho_{(2)}$ remains close to experimental bounds instead of decoupling
like $\Delta\rho_{(1)}$.  Their derivation treats strongly interacting physics
naively, but should give trustworthy orders of magnitude.

   In order to suppress this second order contribution to $\Delta\rho_{(2)}$,
ETC models in which $m_t$ is generated by a strong ``top condensate''
self-interaction rather than the usual ``sideways'' interaction have
been proposed \cite{topcondensate}.
In those models the strong interaction that generates the top
self-interaction must be tuned to keep $m_t$ at the weak scale.
However the weak-isospin-violating interactions, acting only on the
$t$ quark, not the technifermions, then suppress the above $\Delta\rho$
contributions by at least $(m_t/\Lambda_{TC})^2$.

   Returning to our original examples of chiral ETC, we lastly consider
corrections to the tree level left- and right-handed $Zbb$ couplings,
$\zeta_{L,R} \equiv (e/s_\theta c_\theta)\,(I_3^{(L,R)} + (1/3) s_\theta^2)$.
These correct the ratio $R_b \equiv \Gamma_{Zbb} / \Gamma_{\rm hadrons}$:
\begin{equation} \label{Rbshift}
 {\Delta R_b \over R_b} \ \simeq\ 2(1 - R_b)
 \left( {\Delta \zeta_L \over \zeta_L} + {\zeta_R^2 \over \zeta_L^2}
 {\Delta \zeta_R \over \zeta_R}\right)
 \left[ 1 + {\zeta_R^2 \over \zeta_L^2} \right]^{-1} \ ,
\end{equation}
which simplifies upon noting that the standard model gives
$\zeta_R^2 / \zeta_L^2 \approx 0.035$, and that in any case
$\Delta\zeta_R$ is negligible in most ETC models.
The adjoint gauge bosons couple only to the technifermions and hence
do not contribute to this process at lowest order.
As shown in Figs.~\ref{fig2}b and \ref{fig2}c, non-vanishing lowest order
contributions come from ETC bosons that couple to both the third
family and the technifamily:  fundamental representation gauge bosons
\cite{zbbtheory}, and singlets \cite{diagETCrho}.
The fundamental (``sideways'') bosons correct the tree level $\zeta_L$ by
\begin{equation}
 \Delta \zeta_L^{\rm fund}\ \simeq\ {e \over s_\theta c_\theta}
 {v^2 \over N_D M^2_{\rm fund}} {l_{\rm fund}^2 \over 4} \ ,
\end{equation}
where (dropping the order unity mixing angles for conciseness)
$l_{\rm fund} \approx g_L/\sqrt{2}$ is the coupling of the $b_L$ quark
to the ETC-flavor-raising gauge boson.
Thus the fundamental contribution is of order
 $\Delta\zeta_L^{\rm fund} / \zeta_L \simeq
  -1.8\%\ (1\ {\rm TeV} / M_{\rm fund})^2$.

   For models possessing broken diagonal generators of the ETC group,
the associated (technicolor-singlet) ETC bosons generate
\begin{equation}
 \Delta\zeta_L^{\rm diag}\ \simeq\
 {e \over 2 s_\theta c_\theta} {v^2 \over N_D M_{\rm diag}^2}
 \left[ A^{\rm diag}_U - A^{\rm diag}_D \right] l_{\rm diag} \ .
\end{equation}
Here, $l_{\rm diag} \approx -g_L$ up to mixing angles, reflecting the sign
difference between the (traceless) diagonal generator's $b$ and $D$ entries.
With $A_U^{\rm diag} \gg A_D^{\rm diag}$, again reflecting the $t-b$ mass
splitting, this shifts $\zeta_L$ in the opposite direction,
 $\Delta \zeta_L^{\rm diag} / \zeta_L \simeq
 +7.4\%\ N_{TC}^{-1}\ (1\ {\rm TeV} / M_{\rm diag})^2$.
The factor of $1/N_{TC}$, resulting from the gauge generator normalization
($1/\sqrt{(N_{TC}(N_{TC}+1)}$ for the fundamental representation of
$SU(N_{TC})$, acts to suppress this diagonal contribution; nevertheless
for relatively small $N_{TC}$ it may exceed the fundamental contribution.
It has been suggested \cite{diagETCrho} that this positive diagonal
contribution is responsible for the observed excess \cite{lepwg} in the
$Zb{\bar b}$ branching ratio.
However, we have argued that to be compatible with the $\rho$ parameter
data the lowest ETC scale must be at least of order 10 TeV;
then these $Zb{\bar b}$ vertex contributions are also suppressed by two
orders of magnitude.  Such effects are unobservable in current experiments.
These contributions are also suppressed in ETC models which generate the top
mass via a strong $t$ quark self-interaction; the sideways and fundamental
ETC couplings are weak since they are not responsible for $m_t$.

   In conclusion, we have observed that realistic ETC models are chiral.
Models with separate chiral groups, which we have argued are most natural,
contain massive ETC bosons in the technicolor adjoint representation.
In models that generate $m_t$ perturbatively, these gauge bosons give
rise to a contribution to $\Delta\rho$ of about 6\%, more than an
order of magnitude above the experimental limits.
Although that contribution may be suppressed, by raising the lowest ETC
scale and tuning the ETC couplings to generate the large top mass,
corrections to the $Zb{\bar b}$ vertex are then also suppressed below
observable limits.
Model building must either accept such implications, or construct a
framework that explicitly avoids introducing adjoint gauge bosons.
In either case it is essential to consider complete models of both $t$
and $b$ mass generation.

\vfil

   We thank John Terning and Sekhar Chivukula for useful discussions and
criticism.  This work was supported in part under U.S. Department of Energy
contract No.\ DE-AC02-ERU3075.

\end{document}